\documentclass{raa}
\usepackage{graphicx,times,natbib,amssymb,amsmath,url,float,longtable,booktabs,array,calc}
\bibpunct{(}{)}{;}{a}{}{,}

\renewcommand{\del}[1]{}

\title{SVOM Ground Support System}
\author{
  Liu Yurong\inst{1}
  \and Bai Meng\inst{1,2 \star}
  \and Wei Mingyue\inst{1}
  \and Feng Ke\inst{3}
  \and Li Boquan\inst{1}
  \and Hu Tai\inst{1}
}

\institute{
  National Space Science Center, Chinese Academy of Sciences \\
  \and
  University of Chinese Academy of Sciences \\
  \and
  Aerospace Information Research Institute
}

\abstract{
The Ground Support System (GSS) is a critical component of the
Space-based multi-band astronomical Variable Objects Monitor (SVOM)
mission's ground segment. Its main tasks include organizing and
implementing the operations and management of the SVOM payloads,
receiving scientific data, as well as preprocessing and managing the
scientific data. 
To address the specific requirements of the SVOM mission, including specialized mission planning, data reception, status monitoring, and product processing, a dedicated SVOM Ground Support System has been designed based on the integrated architecture of the Chinese Space Science Satellite Project Ground Support System. This system has successfully supported SVOM's launch and on-orbit operations.
}

\begin{document}

\maketitle

\footnotetext{$*$ Corresponding authors: these authors contributed equally to this work.}


\section{Introduction}\label{sect:intro}

SVOM mission is a collaborative mission between China and France aimed at detecting,
localizing, and studying Gamma Ray Bursts (GRB) and other high-energy
transient phenomena (\citep{Cordier+etal+2026b}).

\begin{figure}[htbp]
  \centering
  \includegraphics[width=\linewidth]{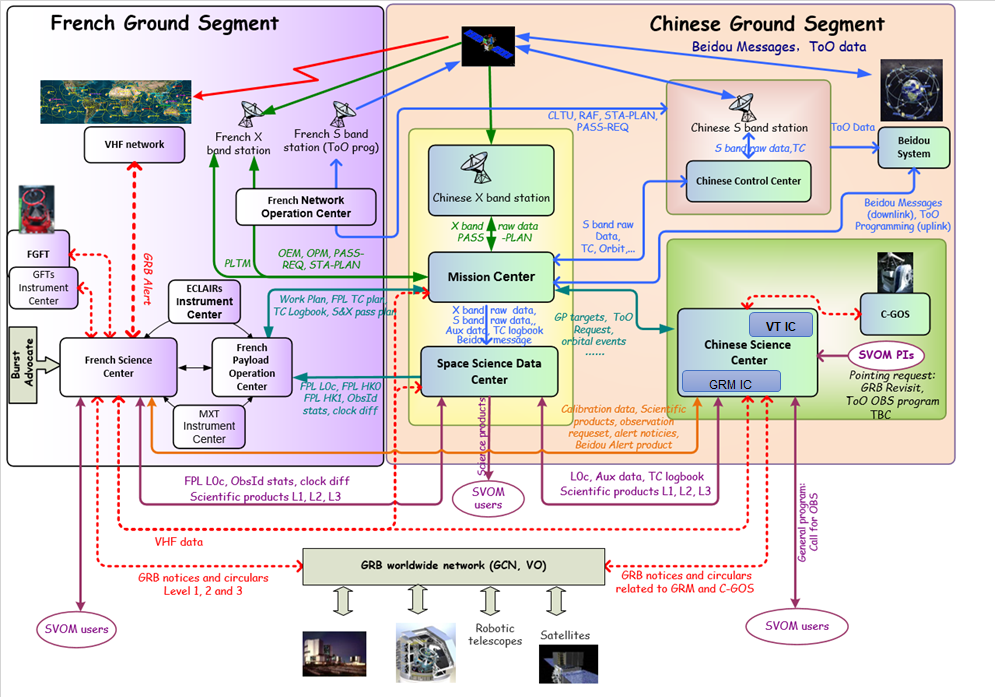}
  \caption{Overall of SVOM Mission.}
  \label{fig:mission}
\end{figure}

SVOM mission is a complex space-ground integrated system composed of
on-orbit satellite and ground segments, including the Chinese and French ground segments, as shown in Figure~\ref{fig:mission}.
The GSS is a component of the Chinese ground segment of SVOM and is the
core of SVOM operations. It acts as a hub connecting satellite and other
centers, achieving operational control and management of SVOM payloads,
scientific data reception, as well as preprocessing and management of
scientific data (\citep{cordiera+etal+2026}).

In addition, the GSS provides dedicated service tools to support
scientific users for efficient scientific research and applications. These tools include observation status query services, Operation Coordination Group (OCG) support services, and data services, among others.

\subsection{System Requirements}

The SVOM GSS consists of three main components: the Mission Center (MC), the
Space Science Data Center (SSDC), and the ground receiving subsystem,
which includes a Chinese X band ground station.

\subsection{Requirements of Mission Center}

The Mission Center is the coordination and operation center for SVOM
mission, serving as the in-orbit management center for the payload. Its
main tasks include mission planning and uplink control, along with
downlink data processing and real-time monitoring.

For mission planning and uplink control, MC is responsible for
generating work plans, integrating routine observations, opportunity
observations, gamma-ray burst observations. It conducts comprehensive planning and analysis of observation tasks and distributes the results. It is responsible for generating
control commands for Chinese payloads, collecting and merging control
commands from both Chinese and French sides, and it plans the data
transmission schedule for the onboard X-band transmitter, generating
command sequence files, and sending them to the tracking and control
system. For ToO urgent observations, immediate uplink can be
performed via the BeiDou system.

For downlink data processing and real-time monitoring, MC is responsible
for collecting S-band and X-band downlink data sent by ground stations
and promptly sending the data to the SSDC; it processes downlink data in
real time, monitors and displays payload status parameters,
and tracks key parameters related to satellite operation conditions. Moreover, MC sends the monitoring results to the Chinese Science Center for
on-board payload status assessment, and is also responsible for the
in-orbit status management of the payload.

\subsection{Requirements of Space Science Data Center}

The Space Science Data Center is the data processing, management, and
archiving center for SVOM mission. Its main tasks include data
preprocessing and product distribution, along with data archiving and
release.

SSDC in charge of preprocessing and performing quick-look transformation of satellite
raw data to generate primary scientific data products and
auxiliary data products. These products are
then distributed to the scientific centers of both China and France, and
the payload operation center in France receives the payload engineering
data products.

Furthermore, the center is responsible for collecting, uniformly managing, and
long-term preserving all levels of scientific data products, auxiliary
data products, and document tools generated by SVOM mission. It creates
scientific data archives and catalogs to ensure the completeness,
security, and permanent usability of scientific data, providing data
release services to scientific users.

\subsection{Requirements of the ground receiving subsystem}

The ground receiving subsystem provides data reception by the Sanya
station, with a reception time exceeding four tracks per day. The
received data is promptly transmitted via the communication network to
the Mission Center for further processing.

\subsection{SVOM Mission Special Requirements}

Compared with general scientific satellites, the SVOM mission has the following special requirements, which need to be focused on in system design.

\begin{enumerate}
  \renewcommand{\labelenumi}{\arabic{enumi})}
  \item Mission planning and control integrating the General Program (GP) and the Target of Opportunity Program (ToO)
  
SVOM’s on-orbit observations are categorized into three types: the Core Program (CP) driven by on-orbit detections, the GP targeting pre-selected sources, and the ToO program responding to ground requests for active sources. For GP and ToO observations, the MC is required to conduct planning and command uplinking based on the scientific teams’ source catalogs and observation requests—demanding high planning efficiency and rapid uplinking, which is a key design focus of the MC.

  \item Reception, processing, and monitoring of multi-source data
 
For data downlink, SVOM transmits data through multiple channels: S-band telemetry (for status monitoring), X-band data transmission (primary scientific data), VHF channel data (scientific and engineering), and Beidou short message data (scientific and engineering). Real-time processing and monitoring of these multi-source data are critical functions of the MC, while dedicated data product generation is the core responsibility of the SSDC and requires SVOM-specific design.

\item Complete and reliable reception of scientific data

As the main channel for scientific data, the X-band relies on collaboration between Chinese and French ground stations, the MC, and the SSDC to ensure downlink integrity and timeliness: ground stations transmit received data to the MC in near real time; the SSDC assesses data quality during product generation; and the MC arranges supplementary downlink for missing data when necessary, thus guaranteeing data integrity.
\end{enumerate}

\section{The Design of SVOM Ground Support System}

\begin{figure}[htbp]  
    \centering  
    \includegraphics[width=0.95\columnwidth]{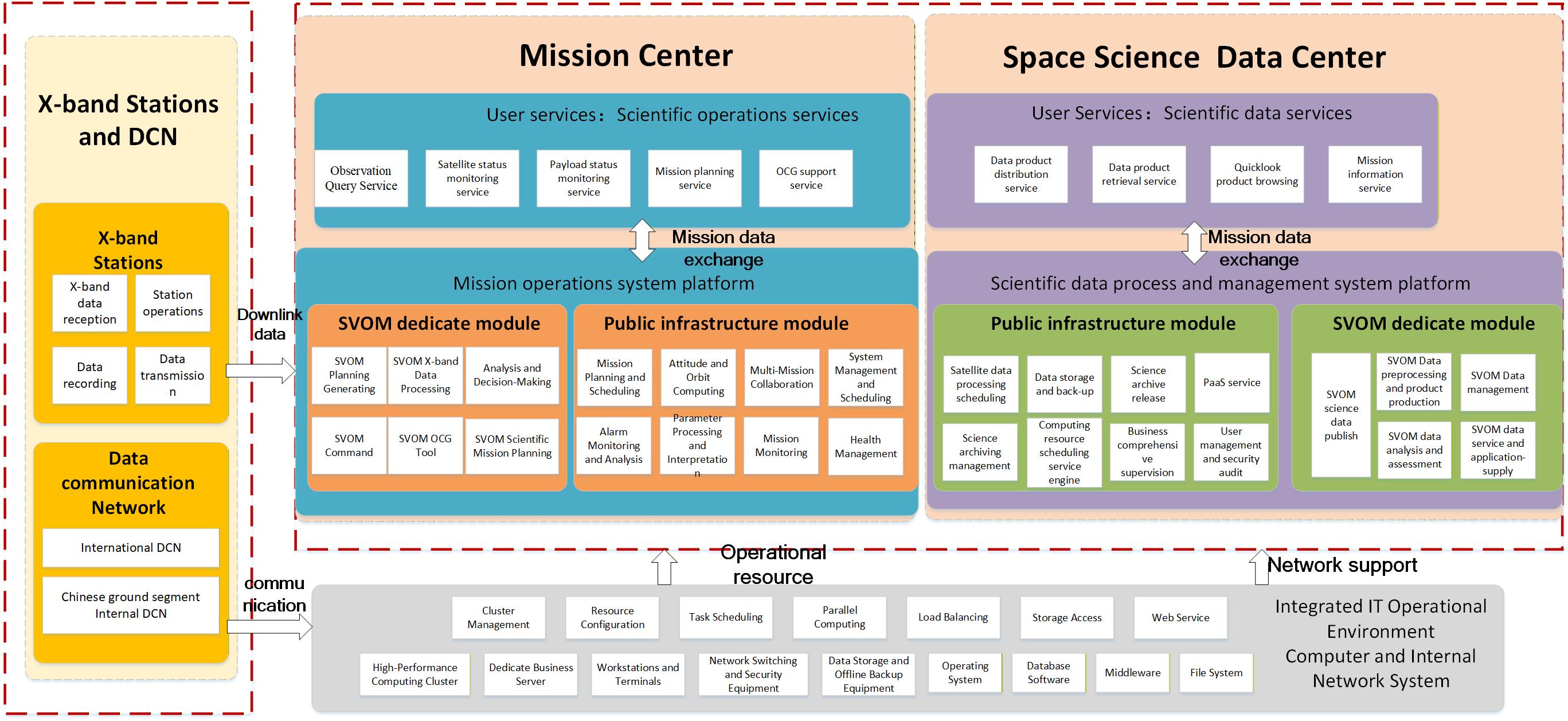}  
    \caption{The architecture of SVOM Ground Support System}  
    \label{fig:myfig}  
\end{figure}

The SVOM GSS is designed based on the overall architecture of the Chinese Space Science Satellite Project Ground Support System, adopting a combination of public infrastructure modules and SVOM-dedicated modules to support the SVOM mission, as illustrated in Figure~\ref{fig:myfig}. 

The Chinese Space Science Satellite Project Ground Support System provides system operation resources and network support for the MC and SSDC through the integrated IT operational environment computer and internal network system. Both the MC and SSDC are internally divided into public infrastructure modules and SVOM-dedicated modules: the public infrastructure modules can support multiple satellite missions, while the dedicated modules are customized for the SVOM mission. The SVOM-dedicated modules of the MC include planning generation, commanding, X-band data processing, OCG tool, analysis and decision-making, and scientific mission planning. The SSDC’s dedicated modules consist of data pre-processing and product production, data management, data analysis and assessment, data service and application provision, and scientific data publication. The MC and SSDC provide mission operation and data product services to scientific users through user service modules. 
The Data Communication Network (DCN) of the GSS is also multi-mission-capable, including the internal communication network of the Chinese ground segment and the international communication network. It supports data transmission from ground stations to the MC and SSDC, data exchange between the ground support system and the Chinese Control Center (CCC) as well as Chinese Scientific Center (CSC), and data transmission with the French Scientific Center (FSC), French Payload Operations Center (FPOC), and Network Operations Center (NOC) via the international DCN.
  For the SVOM mission, the Sanya ground station of the ground support system is utilized, and together with the French X-band stations, completes the reception of SVOM X-band downlink data.

\section{Mission Center}
\label{sect:MissionCenter}

\subsection{Mission planning and payload controlling}

There are two types of SVOM mission planning: GP planning and ToO planning. MC is responsible
for scientific observation scheduling and TC generation.

\begin{enumerate}
\renewcommand{\labelenumi}{\roman{enumi}.}
\item \textbf{GP Programming}

  \begin{figure}[htbp]
    \centering
    \includegraphics[width=\linewidth]{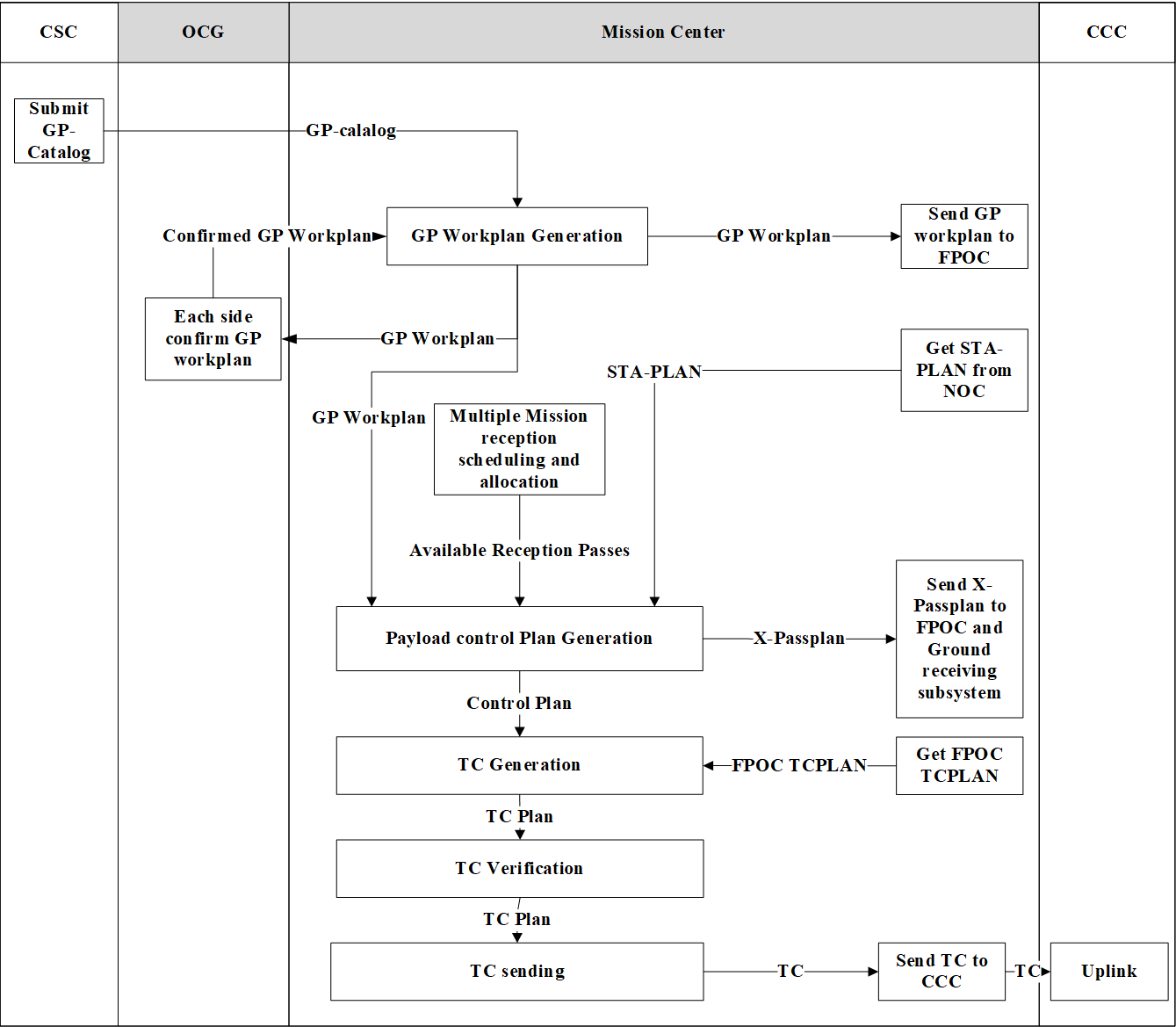}
    \caption{GP Planning in MC.}
    \label{fig:mission2}
  \end{figure}

  The GP planning (as shown in Figure~\ref{fig:mission2}) consists in updating, on a weekly basis, the general program for the following 7 days. The Mission Center
  splits the observing program into 7 days Work Plan.

  Each week, the MC implementation flow is as follows:

  \begin{enumerate}
  \renewcommand{\labelenumii}{\arabic{enumii})}
  \item MC receives the GP-Catalog which is submitted by CSC in advance.
  \item MC conducts mission scheduling and generates the GP-Workplan.
  \item The OCG decides the 7-day Work Plan, using the GP Work Plan generated by MC as input based on the Orbital
        Ephemeris Message (OEM). The program also includes nominal calibrations.
  \item MC sends the GP telecommand plan (including pointing plan, general
        slew points, earth eclipse, etc.), Chinese payload and French payload
        telecommand plans to CCC.
  \item CCC collects the 7-day Chinese and French payload telecommand plan and
        the satellite platform telecommand plan, uplinks them on available
        passes before execution.
  \end{enumerate}

\item \textbf{ToO Programming}

\end{enumerate}

\begin{figure}[htbp]
  \centering
  \includegraphics[width=\linewidth]{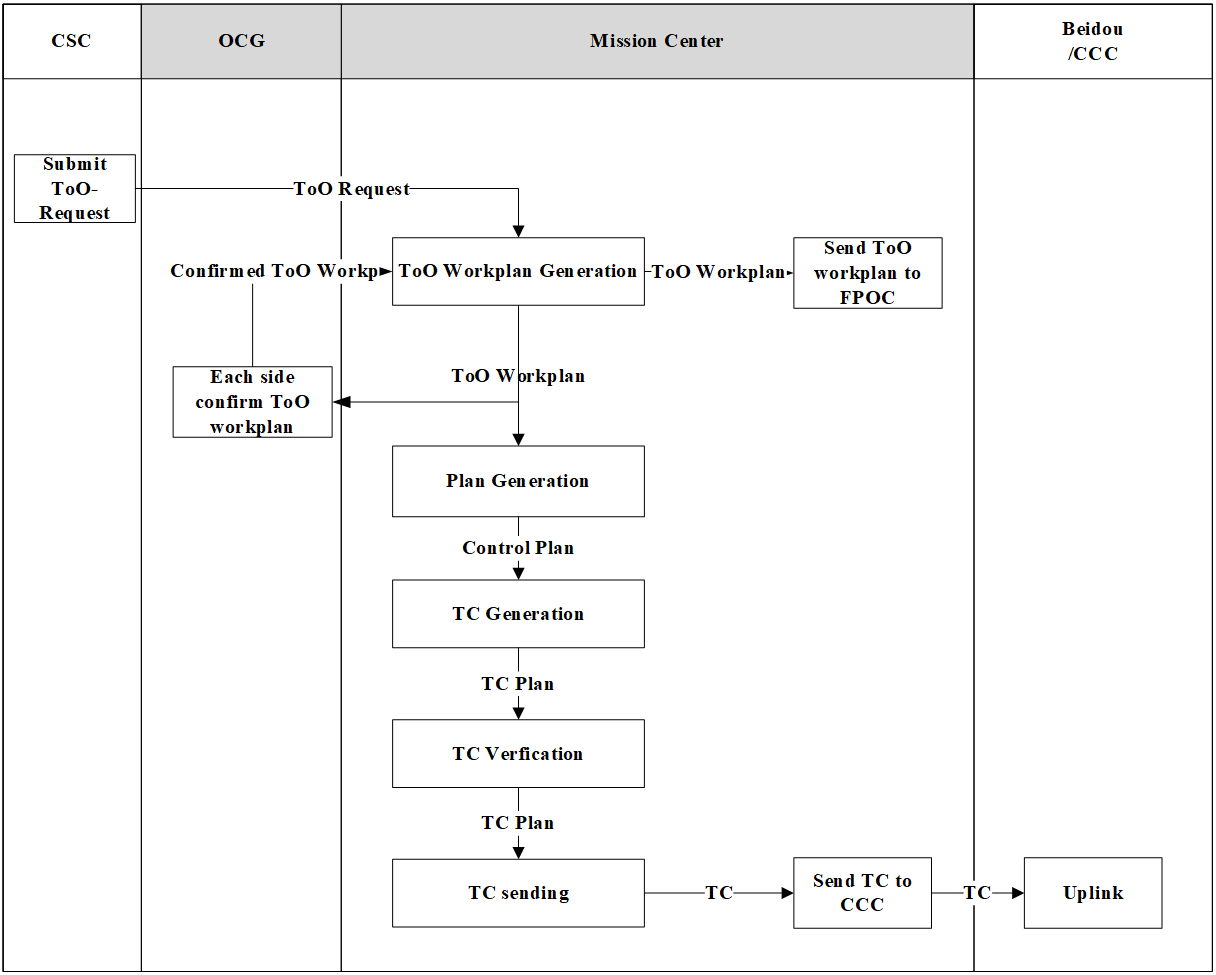}
  \caption{ToO Planning in MC.}
  \label{fig:mission3}
\end{figure}

The ToO planning process (as shown in Figure~\ref{fig:mission3}) has to be able to
modify the global observation timeline to take into account urgent
observing needs (including GRB revisits) that are submitted with ToO
requests (Nominal or Exceptional or Multi-Messenger ToO). These requests
have to be agreed by the PI or a ToO scientist and are elaborated by the
CSC and then, transmitted to the MC. Furthermore, non-routine
calibration needs will also be fulfilled by these asynchronous ToO
requests.

Nominal ToO shall be started within 48 hours from ToO acceptance or
triggered by the ToO Scientist and PI.

Exceptional ToO shall be started within 12 hours (goal) from acceptance
by the ToO Scientist and PI. The default uplink channel to satellite is
Beidou short message.

Multi-messenger ToO shall be started within 12 hours (goal) from
acceptance by the ToO Scientist and PI. The default uplink channel to
satellite is Beidou short message.

During Operation phase, above 80\% percent of the ToO-EX and ToO-MM is
uplink by Beidou short message, the time delay is less than 1 hour (\citep{Bai+etal+2026}).

Once PIs agree on the ToO proposal, the ToO request is then sent to MC, which then engages the ToO programming process depending on ToO category as
mentioned above.

\subsection{Mission monitoring}

MC is monitoring the SVOM mission key parameters and health checking by
S-band, X-band, VHF band and Beidou downlink data. MC also provides the
monitoring Service for the team.

\begin{enumerate}
\renewcommand{\labelenumi}{\roman{enumi}.}
\item \textbf{Real-Time Multi-Channel Data Processing}
\end{enumerate}

MC has implemented an integrated downlink processing system that
seamlessly combines data from multiple channels, including Beidou,
Telemetry, Tracking \& Command (TT\&C), and VHF channel. By leveraging
advanced fusion algorithms and parallel processing technologies, the
system achieves real-time data integration and decoding. This ensures
timely, synchronized, and comprehensive handling of heterogeneous data
streams, greatly enhancing the efficiency and reliability of downlink
operations.

\begin{enumerate}
\renewcommand{\labelenumi}{\roman{enumi}.}
\setcounter{enumi}{1}
\item \textbf{User-Oriented Satellite Status Monitoring Service}
\end{enumerate}

A dedicated satellite status monitoring platform has been developed to
provide users with intuitive and detailed visibility into the health and
performance of both the satellite platform and its payloads. This
service features customizable dashboards, anomaly detection alerts, and
trend analysis tools, enabling users to quickly assess system status,
diagnose issues, and support decision-making processes. It has become an
essential tool for ensuring mission status monitoring and maximizing the
value of satellite data.

\section{Ground receiving subsystem}
\label{sect:receivingsubsystem}

\subsection{Introduction to RSGS}

China Remote Sensing Satellite Ground Station (RSGS) was founded in
1986, and now belongs to Aerospace Information Research Institute (AIR),
Chinese Academy of Sciences. The RSGS is mainly responsible for the data
reception task of the SVOM satellite. RSGS has formed an operation
pattern centered around Beijing headquarters for operation
management. This network includes the Miyun, Kashi, Sanya, Lijiang, and Mohe stations, which collectively form the data receiving infrastructure. Sanya station
undertakes the X-band data receiving task in the SVOM mission.

\subsection{X-band Ground Station of RSGS}

\begin{figure}[htbp]
  \centering
  \includegraphics[width=\linewidth]{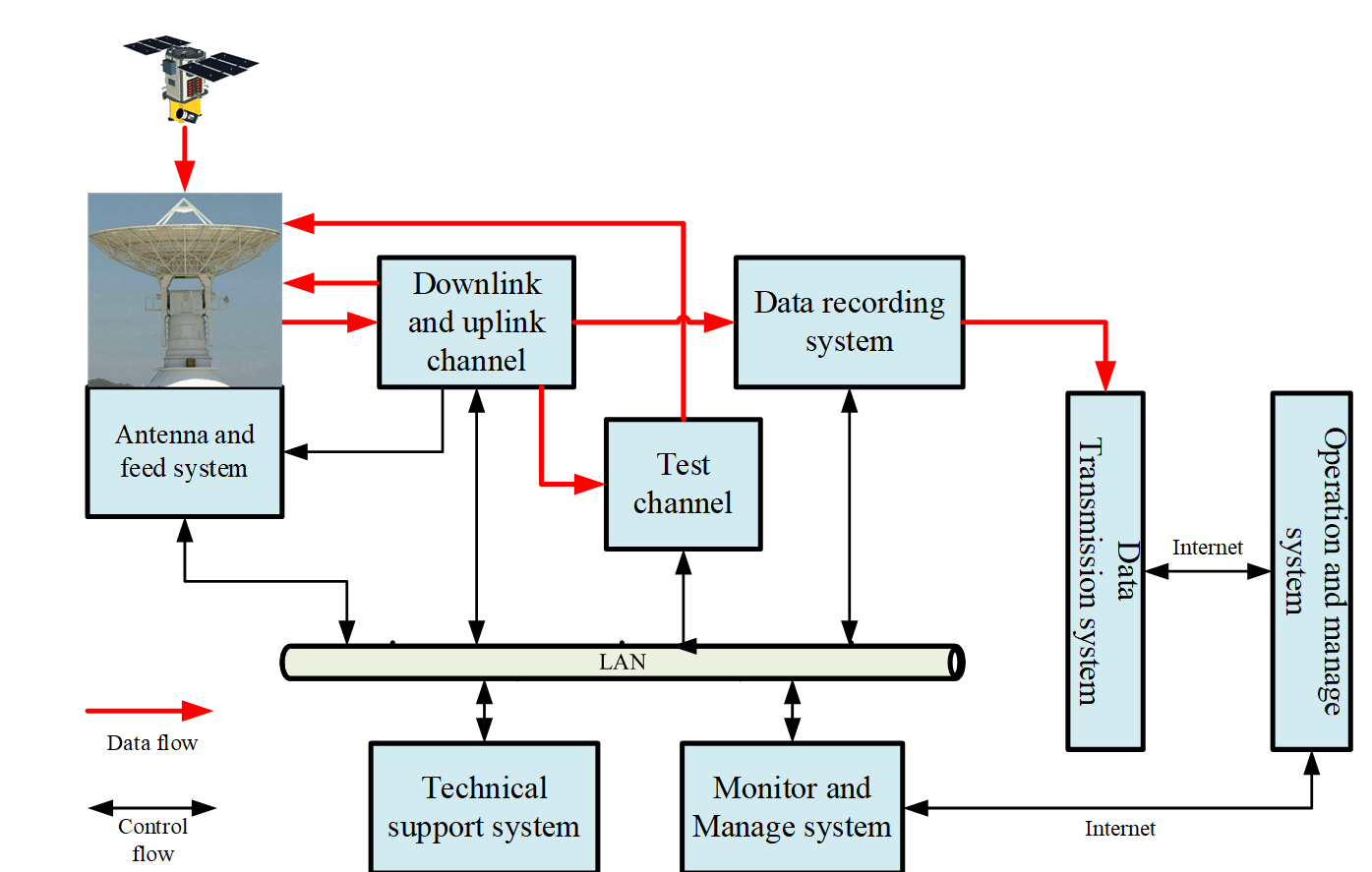}
  \caption{Composition of X band ground station.}
  \label{fig:mission4}
\end{figure}


\begin{table*}[htbp]
  \centering
  \small  
  \caption{Main Characteristics of RSGS X-band ground station.\label{tab:xband}}
  \begin{tabular}{
    >{\raggedright\arraybackslash}p{0.30\textwidth}  
    >{\raggedright\arraybackslash}p{0.65\textwidth}  
  }
    \toprule
    \textbf{Characteristic} & \textbf{Value} \\
    \midrule
    Antenna Size and type & 12-metre diameter cassegrain antenna \\
    Receive RF bands      & S band: 2200\,MHz--2300\,MHz \\
                          & X band: 8025\,MHz--8400\,MHz \\
    Receive G/T           & S band: $\geqslant 20.0\,\text{dB/K}$ (with band-pass filter) \\
                          & X band: $\geqslant 33.5\,\text{dB/K}$ (with radome) \\
    Receive Polarization  & Track: RHC \& LHC selectable in S/X bands \\
                          & Data: RHC \& LHC selectable in S band \\
                          & Data: RHC \& LHC simultaneously in X band \\
    Receive Modulation    & BPSK, QPSK, S/OQPSK, GMSK, 8PSK \\
    Receive Coding        & LDPC, Reed-Solomon, Viterbi \\
    Receive Data Rate     & 500\,K--400\,Msps \\
    Short term record capacity & $\geqslant 10\,\text{T}$ \\
    Monitor and control   & Automatically, Remotely \\
    \bottomrule
  \end{tabular}
\end{table*}

\subsection{Data Flow of Daily Work}

\begin{figure}[htbp]
  \centering
  \includegraphics[width=\linewidth]{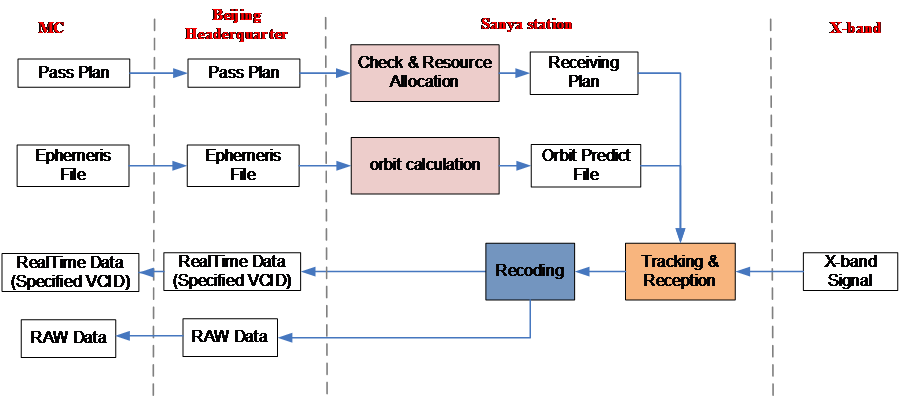}
  \caption{SVOM X band data reception workflow design.}
  \label{fig:mission5}
\end{figure}

The workflow of data reception is as follows:

\begin{enumerate}
\renewcommand{\labelenumi}{\arabic{enumi})}
\item
  Beijing Headquarter of RSGS receives the  pass plan and ephemeris file from
  MC.
\item
  Beijing Headquarter checks if the pass plan and ephemeris file are
  coherent, then send the verified pass plan and ephemeris file to Sanya
  station.
\item
  Sanya station will track, receive, record, transmit the X-band data.
\end{enumerate}

\subsection{Transmission Duration}

Through high-speed network, the average duration of data transmission
for one pass between Station and MC less than 3 minutes in nominal case.

\section{Space Science Data Center}
\label{sect:DataCenter}

\subsection{Science Data Processing Workflow}

\begin{itemize}
\renewcommand{\labelenumi}{\roman{enumi}.}
  \item \textbf{Data product level definition}
\end{itemize}

The SVOM data product are primarily categorized into several
representative classes according to the product level definition, such
as raw data, Level-0, Level-1, Level-2, and Level-3 product. The
specific definitions are summarized as follows:

Raw data: Original downlinked data received by the ground stations,
including four categories: X-band, S-band, VHF, and Beidou.

Level 0 (L0): Binary source packet products generated after data
stitching, sorting, and de-duplication, segmented by observation ID
(OBSID) and hourly intervals.

Level 1 (L1): Products derived from L0 data through source packet
parsing, physical quantity conversion, and formatting. These are
expressed in engineering units (e.g., counts/s).

Level 2 (L2): Calibrated products generated from L1 data by applying
instrument calibration, expressed in physical units (e.g.,
photons/cm²/s).

Level 3 (L3): High-level science products obtained by fusing and
analysing L2 data. These are designed for scientific interpretation,
such as characterizing the variability of physical parameters, applying
model-dependent analysis, or combining data from multiple instruments.

Within this framework, the SSDC is responsible for generating L0 and L1
products, while higher-level science products are produced by the FSC
and CSC (\citep{Louvin+etal+2026}\citep{Huang+etal+2026}).

\begin{itemize}
\renewcommand{\labelenumi}{\roman{enumi}.}
\setcounter{enumi}{1}
  \item \textbf{Data processing workflow}
\end{itemize}

\begin{figure}[htbp]
  \centering
  \includegraphics[width=\linewidth]{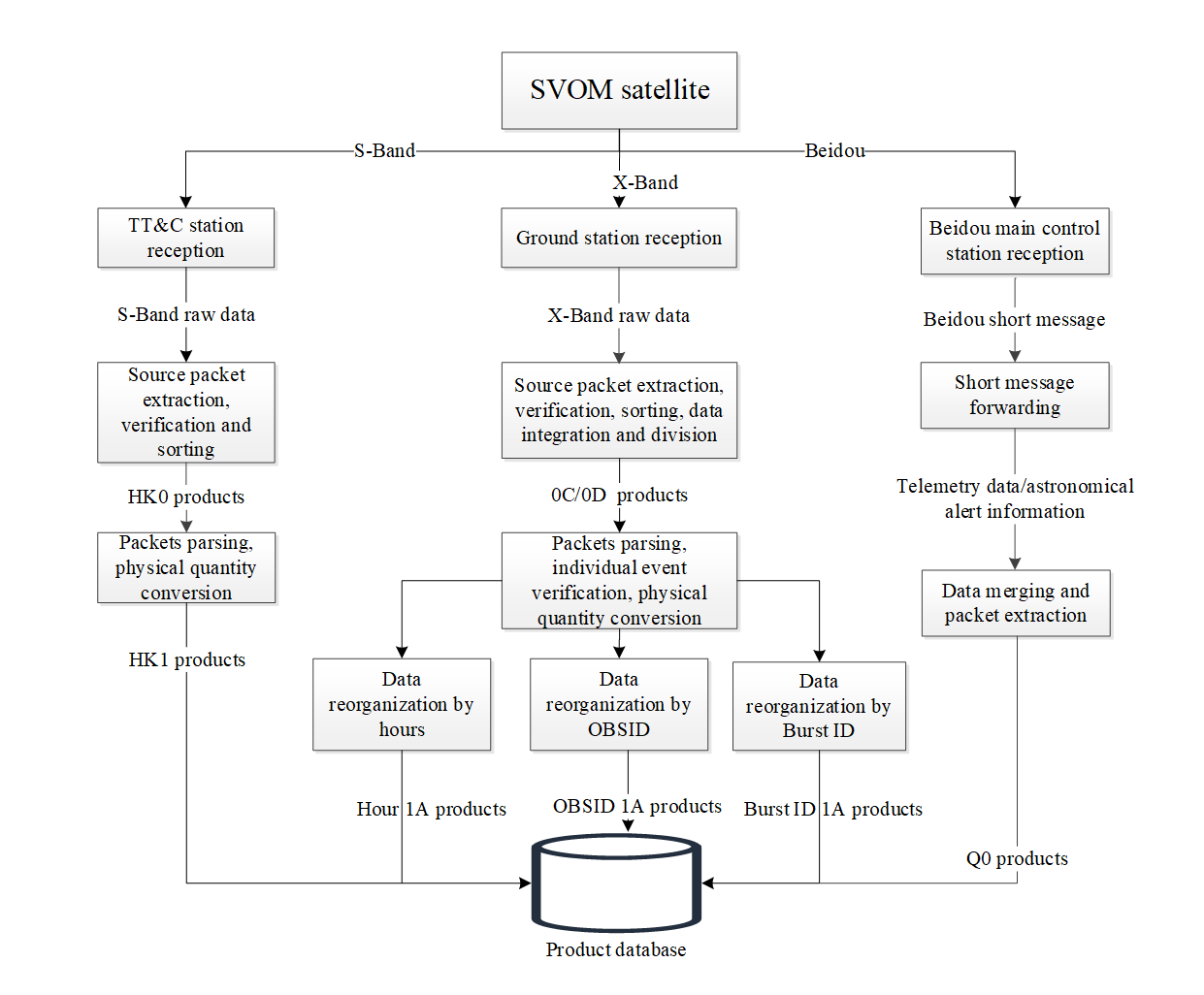}
  \caption{SVOM science data processing workflow design.}
  \label{fig:mission6}
\end{figure}

Based on the hierarchical definition of SVOM data products and their
associated format templates schema, the preprocessing software handles
raw observational data acquired from different downlink channels,
including X-band, S-band, and Beidou. Dedicated data processing plug-ins
are configured separately for each channel according to its specific
requirements, with inter-product dependencies explicitly defined.
Together with job management and resource scheduling mechanisms, this
approach supports the generation of multiple product classes such as
HK0, HK1, L0C, L0D, L1A, and Q0. Each product is accompanied by an XML
metadata file that records quality information. Before ingestion into
the database, all products are automatically validated against
predefined format schema files to ensure compliance with formatting
requirements. The entire data processing workflow is visualized in
Figure~\ref{fig:mission6}.

\subsection{Science Data Product Distribution Workflow}

\begin{figure}[htbp]
  \centering
  \includegraphics[width=\linewidth]{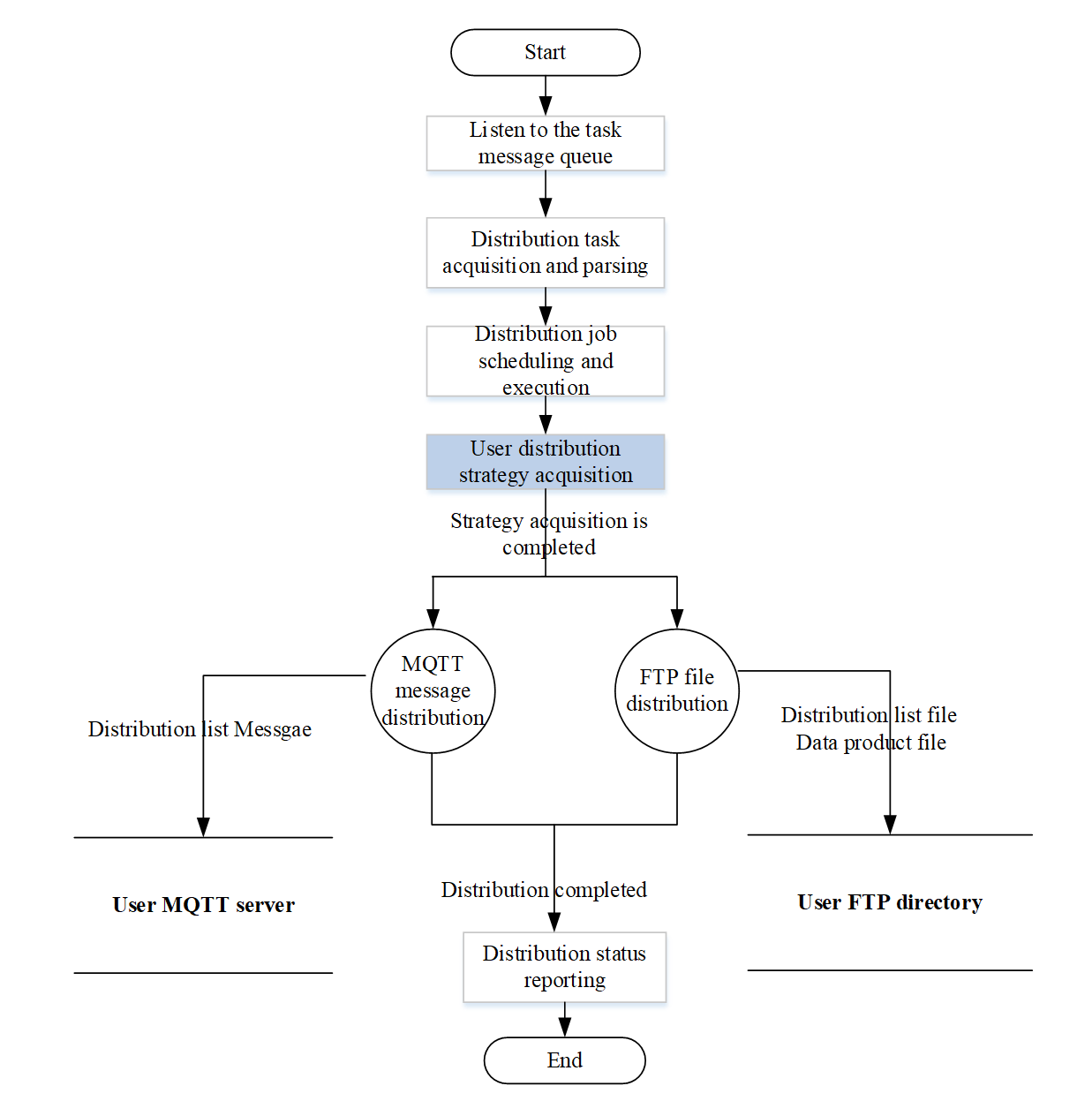}
  \caption{SVOM science data product distribution workflow design.}
  \label{fig:mission7}
\end{figure}

Upon completion of product generation and quality assessment, the
distribution service listens for task notifications via a message queue.
Based on the task list, the system generates distribution jobs that copy
product files into a dedicated cache space. To accommodate the specific
distribution requirements of the SVOM mission, customized strategies are
configured for each product type, covering product categories, target
users, directory rules, distribution lists rules, and priority levels.
Among these, Beidou products are assigned the highest distribution
priority. Following the configured strategies, data products are
delivered to different users (e.g., FSC, CSC, FPOC) via FTP directories,
accompanied by distribution manifest files. Upon successful completion,
the system issues MQTT message to ensure timely distribution
notifications, while scientific team users can monitor distribution
status through the space science mission operation website. The entire
data product distribution processing workflow is visualized in
Figure~\ref{fig:mission7}.

\section{User service}
\label{sect:service}

\subsection{SVOM Observation Query Service}

To efficiently organize SVOM scientific activities in the short term,
scientists need to know when a given target is supposed to
be observed, and when it has been effectively observed. This need arises
from the necessity of coordinating SVOM observations with the schedules
of other observatories, especially those that need to be triggered in a
Target of Opportunity mode. To satisfy the scientific needs, MC has
implemented the SVOM Observation Query (SOQ) tool
(https://soqt.smoc.ac.cn/) where users can search a given observation
either by date, target Id, target name or coordinates, getting the
observation plan and the real execution status. In parallel, the tool
provides the service for searching all the observations performed for a
given source. The tool is very convenient for the scientists to master
the observation status.

\begin{figure}[htbp]
  \centering
  \includegraphics[width=\linewidth]{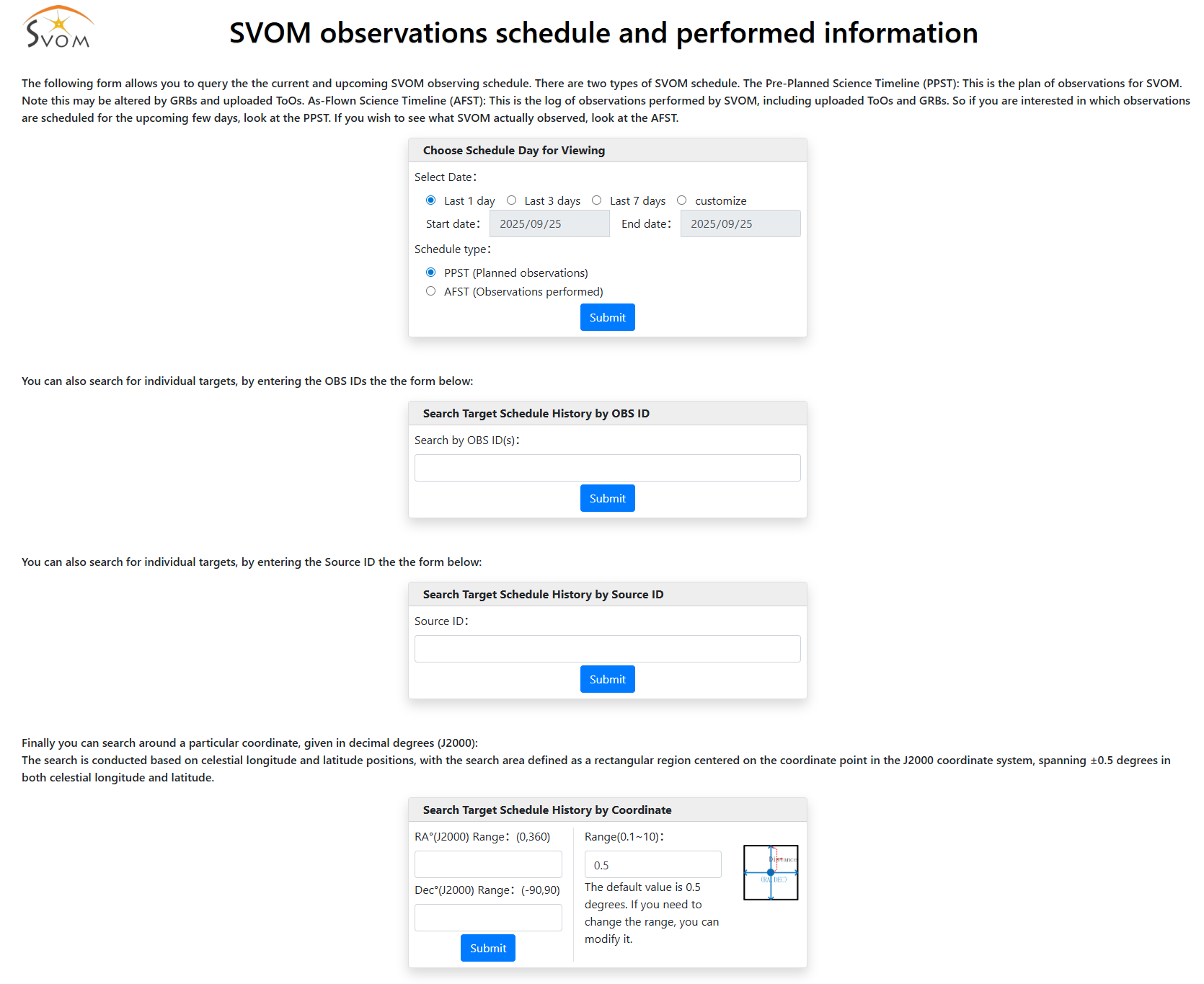}
  \caption{SVOM Observation Query tool.}
  \label{fig:mission8}
\end{figure}

SOQ tool as shown in Figure~\ref{fig:mission8} provides scientists with real-time
visibility into science mission observations, allowing them to
seamlessly track both scheduled and executed observation tasks. It
enables quick comparison between planned and actual observations,
helping scientists identify discrepancies and adapt their strategies.

\subsection{Operation Coordination Group Support Service}

The Operation Coordination Group is a joint working group lead by the
Mission Manager with representatives from all operational teams. It is
the operational entity that validates all operations. It meets regularly
on a daily basis through dedicated system tool (OCG tool
https://ocg.smoc.ac.cn/svom/index), developed by MC. OCG meetings are hold daily during the routine phase to the assess the status of the past weekly operational plan and planification of payload and satellite activities, observations selection for General Program for the next week.

\begin{figure}[htbp]
  \centering
  \includegraphics[width=\linewidth]{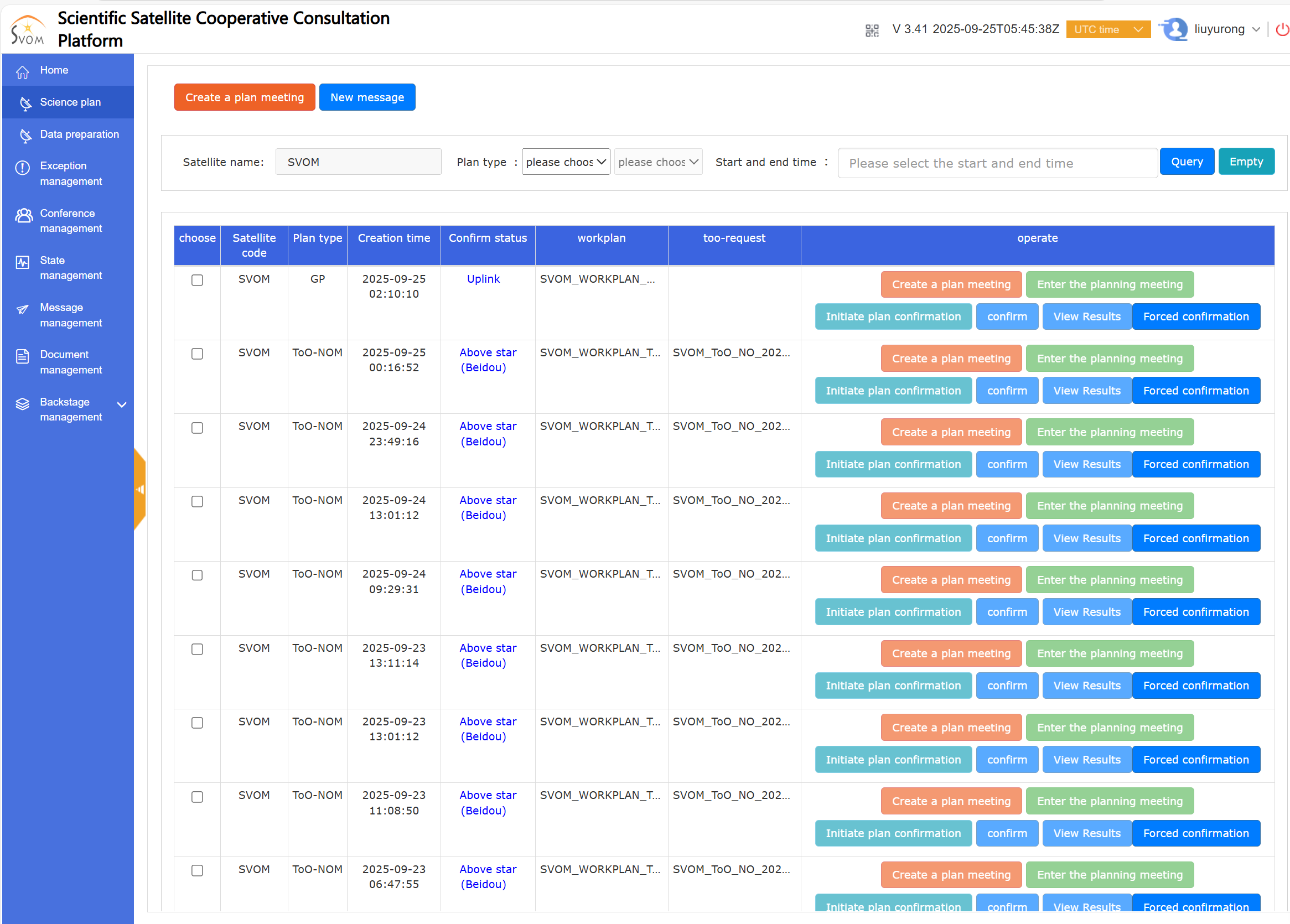}
  \caption{OCG Tool.}
  \label{fig:mission9}
\end{figure}

The OCG tool as shown in Figure~\ref{fig:mission9} aims at helping all operational teams
to make a validated and confirmed plan, through simulation, analysis and
approval procedure. OCG tool primarily focuses on the engineering aspects of the mission rather than the scientific aspects. Periodic status reporting will be
required during operations to provide assessment of the health and
safety of the spacecraft, instruments, and ground systems. These reports are
tackled during OCGs. Daily or weekly status reporting should be
performed to provide all operational teams a global view of SVOM science
mission and engineering status. Exceptional ToO OCG can also be
performed to assess planification of exceptional ToO or Multi-Messenger
ToO. The Figure~\ref{fig:mission9} shows the OCG tool main function.

\subsection{Data Service}

For satellite engineering and science teams, the ground support system
provides rapid-access data services through the space science mission
operation website. These services include near real-time monitoring of
satellite--ground data transmission, status of data product processing
and distribution, data quality inspection, and product overview
visualization. Users can access this information through multiple
interfaces such as web portals and mobile applications, thereby enabling
timely situational awareness of data processing across all product
levels and categories.

\begin{figure}[htbp]
  \centering
  \includegraphics[width=\linewidth]{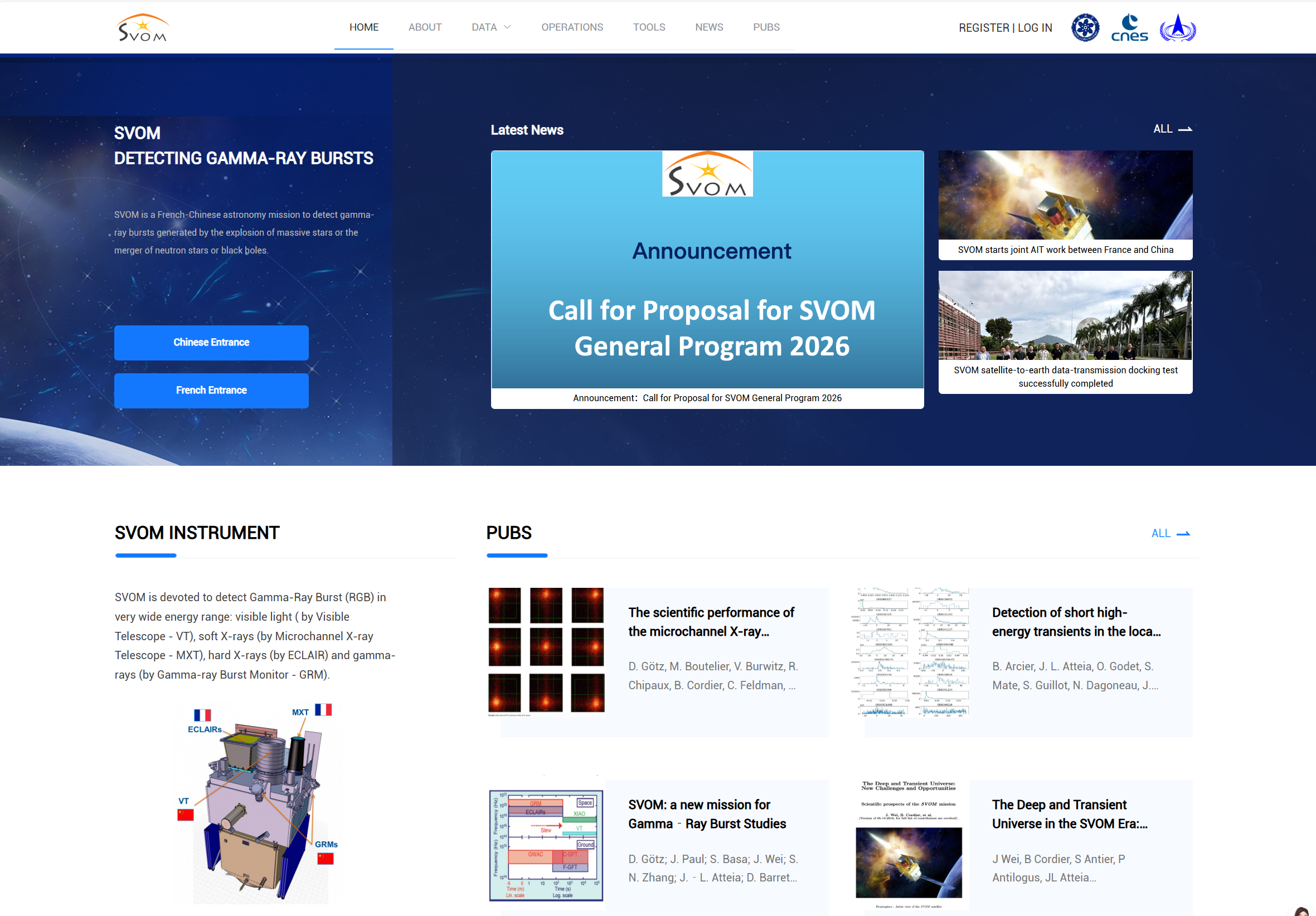}
  \caption{Homepage of the SVOM Satellite Data Release Website.}
  \label{fig:mission10}
\end{figure}

For the public scientific user, SSDC operates a dedicated data release
portal (\url{https://svom.ac.cn}), as depicted in Figure~\ref{fig:mission10}, where users
can search, browse, and download SVOM satellite data, related
documentation, and software tools. The portal supports multi-parameter
search queries based on Burst ID, OBSID, date, source position, payload,
and product level, and also provides quick-look visualizations. In
addition, the portal integrates data policy guidelines, mission
background, and tracking information of scientific achievements.

\section{Commissioning and routine operations}
\label{sect:Operations}

The SVOM satellite was launched on June 22, 2024, and began the commissioning test. During the nine-month Commissioning phase, the MC, SSDC, Sanya X-band station, DCN, and the integrated operational environment operated stably and met all specified requirements.
The MC generated GP work-plans on a weekly basis, which were uplinked to the satellite via the CCC. More than one ToO observation request was uplinked daily, totaling 37 GP work-plans and 492 ToO observation plans—including 54 uplinked through the Beidou system. Currently, one Beidou-uplinked ToO observation is implemented daily. The latency from ToO observation request submission to the start of SVOM satellite observation can be as fast as less than 5 minutes. A total of 1,007 passes of downlink data were received via the Sanya X-band ground station, while 914 passes were jointly received by the French Kourou ground station and Hartebeeshoek ground station. The daily data reception amounts to 8 passes, with a total reception duration exceeding 70 minutes. In accordance with product definitions and interface requirements, SSDC completed the production and distribution of Level 0C, 0D, 1A, Level 1, HK0, and HK1 data products.
The SVOM satellite was officially delivered for on-orbit operations on April 23, 2025, and the SVOM GSS continues to provide support for its on-orbit operations.

During the commissioning phase, the GSS was upgraded to address issues encountered during testing, new requirements, and interface changes. These updates guaranteed data processing accuracy, improved operational efficiency, and delivered better support for scientific users. In response to feedback from the satellite and payload teams on the downlink data monitoring interface and HK products, some parameter processing algorithms were revised in the data processing, mission monitoring, and HK product generation modules. To enhance the rapid uplink success rate of ToO observations, the Beidou short message automatic uplink procedure was optimized with an automatic retransmission mechanism. In response to the new requirement that nominal ToO observations, originally uplinked via CCC, be transmitted with high priority via Beidou short messages, the planning generation, command generation, and Beidou uplink modules were modified. These improvements greatly strengthened the system’s ToO rapid response capability. At the request of the SVOM PIs, an observation status query module (https://soqt.smoc.ac.cn/) and the Chinese SVOM website module (https://svom.ac.cn/) were integrated into the GSS, further enhancing its end-to-end support service capability.

\section{Summary}
\label{sect:Summary}

The SVOM GSS ensures efficient
planning for scientific observation needs, rapid uplink for ToO
observation requirements, complete reception and high-quality processing
and management of scientific data, thus supporting the continuous
production of scientific outcomes for the SVOM mission.
Compared to other space science satellite missions, the SVOM ground support system has achieved 24/7 near-real-time full-system status monitoring, round-the-clock rapid response, and whole-process international cooperation. 
The SVOM ground support system is a comprehensive and distinctive system that can support the SVOM mission, and it also provides a reference for the design of ground support systems for other astronomical satellites. The system can be further improved in user services by drawing on the ground systems of on-orbit astronomical satellites.

\label{lastpage}

\bibliographystyle{raa}   
\bibliography{bibtex}   

\end{document}